\documentclass[journal,transmag]{IEEEtran}

\pdfoutput=1

\usepackage[utf8]{inputenc} 
\usepackage[T1]{fontenc} 
\usepackage{hyperref}  
\usepackage{url}   
\usepackage{booktabs}  
\usepackage{amsfonts}  
\usepackage{nicefrac}  
\usepackage{microtype}  
\usepackage{graphicx}
\usepackage{array}
\usepackage{url}
\usepackage{algorithm2e}
\usepackage{amsmath}
\usepackage{multirow}
\usepackage{float}{
\usepackage{xcolor}
\def\HiLi{\leavevmode\rlap{\hbox to \hsize{\color{yellow!50}\leaders\hrule height .8\baselineskip depth .5ex\hfill}}}
\SetKw{Continue}{continue}

\makeatletter
\newcommand{\removelatexerror}{\let\@latex@error\@gobble}
\makeatother

\newenvironment{conditions}
 {\par\vspace{\abovedisplayskip}\noindent\begin{tabular}{>{$}l<{$} @{${}={}$} l}}
 {\end{tabular}\par\vspace{\belowdisplayskip}}
\def\pshow#1{{\let\protect\show#1}}

\SetKwInput{KwInput}{Input}
\SetKwInput{KwOutput}{Output}

\title{Not all bytes are equal: Neural byte sieve for fuzzing}
\author{
  Mohit Rajpal \\
  Microsoft Research\\
  \texttt{v-mohita@microsoft.com} \\
   \and
   William Blum \\
    Microsoft Research \\
   \texttt{wiblum@microsoft.com} \\
    \and
   Rishabh Singh \\
   Microsoft Research \\
   \texttt{risin@microsoft.com} \\
}

\begin{document}
\maketitle

\begin{abstract}
Fuzzing is a popular dynamic program analysis technique used to find vulnerabilities in complex software. Fuzzing involves presenting a target program with crafted malicious input designed to cause crashes, buffer overflows, memory errors, and exceptions. Crafting malicious inputs in an efficient manner is a difficult open problem and often the best approach to generating such inputs is through applying uniform random mutations to pre-existing valid inputs (seed files). We present a learning technique that uses neural networks to learn patterns in the input files from past fuzzing explorations to guide future fuzzing explorations. In particular, the neural models learn a function to predict good (and bad) locations in input files to perform fuzzing mutations based on the past mutations and corresponding code coverage information. We implement several neural models including LSTMs and sequence-to-sequence models that can encode variable length input files. We incorporate our models in the state-of-the-art AFL (American Fuzzy Lop) fuzzer and show significant improvements in terms of code coverage, unique code paths, and crashes for various input formats including ELF, PNG, PDF, and XML.
%
\end{abstract}

\begin{figure*}[t]
\centering
\removelatexerror
\begin{minipage}{8cm}
  \vspace{0pt}
\begin{algorithm}[H]
\DontPrintSemicolon
\TitleOfAlgo{Simple Random Fuzzing}
\SetKwFunction{RandInt}{RandInt}
\SetKwFunction{len}{len}
\SetKwFunction{mutate}{mutate}
\SetKwFunction{Execute}{Execute}

\KwInput{$Seeds$, Target program $B$}
  \KwResult{$MaliciousInputs$}

\For{$Seed$ $\in$ $Seeds$} {

\For{$iterations \gets 0$ \KwTo $limit$ }{
    $input \gets Seed$

    $length \gets$ \len{$Seed$}

    $mutations \gets$ \RandInt{$length$}

    \For{$mut \gets 0$ \KwTo $mutations$}
    {
        $byte \gets$ \RandInt{$length$}

        \mutate{$input$, $byte$}

    }

   \HiLi $result \gets$ \Execute{$P$, $input$}

   \If {$result$ is crash}
   {
     Append $input$ to $MaliciousInputs$
   }
\HiLi \;
\HiLi \;
\;

 }
}
\end{algorithm}

\end{minipage}%
\begin{minipage}{8cm}
  \vspace{0pt}
\begin{algorithm}[H]
\DontPrintSemicolon
\TitleOfAlgo{AFL Fuzzing}
\SetKwFunction{RandInt}{RandInt}
\SetKwFunction{len}{len}
\SetKwFunction{mutate}{mutate}
\SetKwFunction{Execute}{Execute}
\SetKwFunction{HasInputGain}{HasInputGain}

\KwInput{$Seeds$, Target program $P$}
  \KwResult{$MaliciousInputs$}
\For{$Seed$ $\in$ $Seeds$}
{
 \For{$iterations \gets 0$ \KwTo $limit$ }{
    $input \gets Seed$

    $length \gets$ \len{$Seed$}

    $mutations \gets$ \RandInt{$length$}

    \For{$mut \gets 0$ \KwTo $mutations$}
    {
        $byte \gets$ \RandInt{$length$}

        \mutate{$input$, $byte$}

    }

   \HiLi $result, cov \gets$ \Execute{$P$, $input$}

   \If {$result$ is crash}
   {
     Append $input$ to $MaliciousInputs$
   }
   \HiLi \If {\HasInputGain{cov}}
   {
   \HiLi  Append $input$ to $Seeds$
   }

 }
}
\end{algorithm}

\end{minipage}
\caption{Comparison of the AFL fuzzing algorithm with a simple random fuzzing algorithm. The \emph{mutate} function mutates a byte of the input in place using various techniques such as bit flips, byte flips, bit rotations or arithmetic operations. The \emph{Execute} function executes the target program with the mutated input and reports on crashes. For AFL, the Execute function also reports the induced code coverage ($cov$).}
\label{afl-random-comp}
\end{figure*}

\section{Introduction}

\emph{Fuzz testing}{~\cite{barton90,godefroid12} is one of the most widely used automated software testing technique that has been successful in automatically discovering a large number of security vulnerabilities in complex programs. The key idea in fuzzing is to continuously generate new malicious inputs to stress-test a target program to discover unexpected behaviors such as crashes, buffer overflows, or exceptions. Typically, a \emph{fuzzer} is started with an initial set of \emph{seed} input files, which are continuously transformed to generate malicious inputs either by random mutations, constraint-solving, or using a set of manually-defined heuristics. Since the input formats can be quite complex, generating malicious inputs typically requires millions of mutations, and therefore the fuzzing process can be seen as a huge search problem to identify a good set of mutations that would lead to higher code coverage and more crashes. In this paper, we present a learning technique that uses neural networks to learn patterns in the input files from previous fuzzing explorations to guide the future fuzzing explorations. In particular, the neural models learn a distribution over different locations in the input files to apply the mutations, which in turn is used to guide the fuzzing process to explore new unique code paths and crashes.

The current fuzzing techniques can be broadly categorized into three main categories: i) Blackbox fuzzing~\cite{barton90,radamsa,peach}, ii) Whitebox fuzzing~\cite{godefroid12}, and iii) Greybox fuzzing~\cite{afl}. Blackbox fuzzers treat the target program as a black box with no internal inspection inside the program. In contrast, whitebox fuzzers require knowledge of the structure of the program being tested (possibly, but not necessarily, through analysis of the program source code) to generate input mutations to specifically target certain code fragments. Greybox fuzzers form a middle ground where they perform limited source code inspection such as computing code coverage using lightweight code instrumentation. Although all fuzzing techniques have different strengths and weaknesses, greybox fuzzing techniques based on random mutations have resulted in fuzzers such as AFL (American Fuzzy Lop)~\cite{afl}, which has been successful in finding a large number of real-world bugs in complex programs. The success of greybox fuzzers largely results from their simplicity that allows for efficient low-level implementations.

In this paper, we explore whether it is possible to use machine learning to learn a strategy for guiding the input mutations based on previous history of executed inputs and code coverage information. More specifically, we aim to learn a function that can predict optimal locations in the input files to perform the mutations. We first run the traditional fuzzing techniques for a limited time to obtain data regarding which mutations lead to new code coverage, and then use this data to learn a function to guide further input modifications towards generating new promising inputs. Although our technique is applicable to any fuzzing system, we instantiate it on the current state-of-the-art AFL fuzzer~\cite{afl}, which is a genetic algorithm based greybox fuzzer. AFL performs random mutations to a set of seed input files, and maintains an \emph{input queue} of promising new input files that lead to execution of new code paths. Since it is difficult to precisely mutate the input files using random mutations, typically millions of newly generated inputs are discarded and only a handful of them (in the input queue) are considered for future mutations. Our technique aims to learn to guide this process of input generation to minimize the time spent on generating \emph{unpromising} inputs, thereby increasing the chances of the fuzzer to cover new code paths.

We implemented several neural network architectures to learn the function to predict the expected code coverage map given a set of input modifications. Since input files can be of varying lengths, we use architectures such as LSTM (Long Short term Memory)~\cite{hochreiter97} and sequence-to-sequence with attention~\cite{bahdanau14} that can encode variable-length sequences. At fuzzing time, we use the learnt function to predict a heat map for the complete input file, which corresponds to likelihood of mutations of each file location leading to new code coverage. We then use the coverage map to prioritize the mutation locations. For training these functions, we first run AFL on a subset of seed files for a limited time, and obtain the training data for mutation-coverage information.

We evaluate our technique on several input formats such as ELF, PNG, PDF, and XML. We observe that the neural augmented AFL results in significant code coverage improvements for the ELF and PNG parsers than the AFL, whereas for the PDF and XML parsers, the coverage is comparable. We observe that the neural augmented AFL consistently results in exploring significantly more number of unique code paths for ELF, PNG, and XML parsers. Most importantly, the number of observed crashes significantly increased for neural guided AFL for the ELF and XML parsers. We observe a smaller coverage improvement for PDF parser because of the additional time needed for the learnt model to predict coverage map on large PDF input files, but we believe this performance can be improved through some additional performance engineering.

This paper makes the following key contributions:

\begin{itemize}

\item We model the problem of learning promising locations to fuzz in input files using different neural network architectures.

\item We present a technique to efficiently train the location prediction function and then use the learnt function to perform fuzzing.

\item We implement our learnt neural models inside the state-of-the-art AFL fuzzer, and show that it results in significantly more code coverage, unique code paths, and crashes on different input formats.

\end{itemize}

\section{AFL Background}

AFL is a state-of-the-art greybox evolutionary fuzzer. AFL has a simple strategy to craft malicious inputs: attempt many small localized mutations to the seed files, as well as some \emph{stacking} mutations which mutate many locations in the seed simultaneously. AFL's strength lies in its genetic algorithms. AFL instruments source code during compilation to gain access to code coverage during execution. During execution of the target program, AFL observes the code coverage that a \emph{mutated} seed induces. A mutated seed is considered interesting if it induces some never-seen-before piece of code to be executed, or if it changes the frequency of execution of a previously seen piece of code. This is referred to as \emph{input gain}. AFL saves mutated inputs which induce an input gain and treats them as further seed files to mutate. This constant evolution of the seed pool helps reach many obscure code paths which require many iterated small mutations to reach. This pool is also frequently culled to pick the best seeds to mutate. AFL's strategy has discovered many bugs in mature open source projects such as Mozilla Firefox, ffmpeg, OpenSSL, clang and others. A comparison of a simple blackbox random fuzzer and AFL fuzzer's core algorithm is shown in Figure~\ref{afl-random-comp}.

Fuzz testing is computationally intensive. Even a small input gain requires thousands to millions of random mutations to discover. However, not all mutations are created equal. File formats and their parsers are heterogenous. We believe a mutation to the file header, or other critical sections is more likely to yield input gain. This is a likely scenario since many conditional branches are dependent on small critical sections. In contrast, sections containing raw data are less likely to yield input gain because they are usually read by small pieces of code in tight loops. However, it is difficult to manually identify such locations for a complex input format without a lot of domain expertise.

A natural next step is to codify quantitative techniques to automatically identify optimal locations to mutate. We investigate neural networks based machine learning techniques to automatically identify useful locations in input files using code coverage feedback.

\begin{figure*}[t]
\begin{center}
\includegraphics[width=5in]{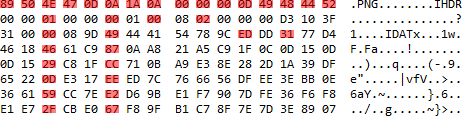}
\end{center}
\caption{A highlighted seed file for the PNG file format. Useful sections are indicated in red. The png header and other relevant bytes are identified as promising locations to mutate.}
\label{png_heatmap}
\end{figure*}

\section{Overview of the Framework}

Our framework consists of a fuzzer and a model that highlights useful locations in an input file. During runtime, the fuzzer queries the model for each seed file and focuses mutations on the highlighted locations. A sample highlighted seed file is presented in Figure \ref{png_heatmap}. Given an input file in a sequence-of-bytes format, the model annotates a heat map function highlighting the relative efficacy of mutating each position in the input file. Due to the variable length of seed files, the model is defined as a family of functions.

\begin{equation}
\left \{f_k: \left \{\text{0x}00, \text{0x}01, \ldots , \text{0x}FF \right \}^k \mapsto [0,1]^k \middle | k \in \mathbb{N} \right \}
\end{equation}

For simplicity, we denote this family of functions as simply $f$ that can take as input an arbitrary number $k$ of input locations. This function associates each position in the input file with the probability of a mutation yielding an input gain. During augmented execution, the fuzzer first queries this model prior to performing mutations, and uses the resultant heatmap to guide the mutation towards useful locations. Potential inputs which target few useful locations are vetoed during augmented execution; this saves time by avoiding executions on inputs which are unlikely to give input gain. Formally, a mutated input is vetoed unless it meets the required cutoff presented in equation \ref{eqn:vetoing}.

\begin{equation}
\label{eqn:vetoing}
\sum_k\left[\left (x \oplus x' \right ) \land \left \lceil f(x) \right \rceil \right] > \alpha
\end{equation}
where:
\begin{conditions}
 x   & input file to fuzz \\
 x'   & mutated input \\
 $$\oplus$$ & bitwise exclusive-or \\
 $$\lceil \_ \rceil$$ & ceiling function \\
$$\alpha$$ & A user defined cutoff hyperparameter
\end{conditions}

Ostensibly, $(x \oplus x')$ is the \emph{diff} of the mutated input with respect to the seed. The key idea in equation \ref{eqn:vetoing} is to consider only diffs which modify many useful byte positions as indicated by $f(x)$. The $\alpha$ parameter controls how many "useful" bytes must be mutated. This formulation of heat map function $f$ is easy and efficient to integrate with any fuzzing system as it performs the heat map computation once in the beginning for any seed file.

For training a model to learn the function $f$, the input file and corresponding code coverage are required. In particular, the following elements are used to train the model:

\begin{itemize}
    \item $x$: The seed file being fuzzed;
    \item $b$: Code coverage bitmap yielded by executing the target program on $x$;
    \item $x'$: The mutated seed file;
    \item $b'$: Code coverage bitmap yielded by executing the target program on $x'$.
\end{itemize}

Note that these data elements are first class citizens of most greybox fuzzers and require no additional instrumentation to generate. Blackbox fuzzers can also be easily augmented to generate code coverage information for the target program.

While it is clear that a lack of change in code coverage indicates mutations applied to \emph{useless} locations, there is no straightforward method to determine \emph{useful} locations through (input, code coverage) tuples. A general framework for creating a supervised training dataset pairs on some scoring of $b, b'$, denoted $s(b,b')$, is:

\begin{figure*}[t]
\begin{center}
\includegraphics[scale=1.2]{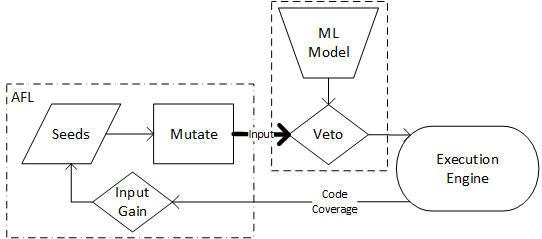}
\end{center}
\caption{The veto model used by Augmented-AFL. Mutations which target no useful locations are not executed.}
\label{augmented_afl_plant}
\end{figure*}

\begin{equation}
{\cal{X}}{\cal{Y}} \> = \> \left \{\left (x, x \oplus x' \right ) \> \middle | s \left (b,b' \right ) > \gamma \right \}
\end{equation}

for some real valued cutoff value $\gamma$. Given the training dataset, the goal is to learn a model that can map an input file $x$ to a diff heatmap $x\oplus x'$, which in turn can be used to identify potentially useful locations to focus the attention of fuzzing mutations.

The advantage of the above approach is that it results in culling the useless mutations which are scored lower than the useful mutations. A model learning from the dataset will receive many well scored $(x, x \oplus x')$ pairs in a supervised setting. A single seed is typically paired with many mutations. To minimize the aggregate loss over this "one-to-many" relationship, the expected value of the diff $x \oplus x'$ given $x$: $E[(x \oplus x') | x]$ is learned. This captures the relative usefulness of flipping bytes at certain locations.

In principal, an effective incarnation of $s(b,b')$ is challenging. The desired behavior of $s(b,b')$ is highlighting mutations which cause input gain, resulting in "never-seen-before" execution behavior in the target program. This sequential dependency on fuzzing history requires a function $s^*$ that is conditioned on previous coverage history, which is unfortunately difficult for learning methods to model. We, therefore, choose an intuitive approximation of $s^*$:

\begin{equation}
  s(b,b') = \sum_{1 \leq i \leq |b|} [b_i < b'_i]
\end{equation}
where $b_i$ denotes the $i$th bit of the bitmap $b$ and $|b|$ denotes the length of the bitmap.
The truth table for bitwise strictly less than function is given in Table~\ref{tabbitwiseless}.

\begin{table}[ht]
\centering
\caption{The truth table for bitwise strictly less function.}
\label{tabbitwiseless}
\begin{tabular}{|l|l||l|}
\hline
x & y & $x < y$ \\ \hline \hline
0 & 0 & 0                  \\ \hline
0 & 1 & 1                  \\ \hline
1 & 0 & 0                  \\ \hline
1 & 1 & 0                  \\ \hline
\end{tabular}
\end{table}


The bitwise `strictly less than' scoring function highlights code sections which are not executed in $b$, but executed in $b'$. This function rewards increases in code coverage. In practice, we have found this scoring function to give good results across a wide number of target program.

\section{Learning for Augmented Fuzzing}
Our design for learning augmented fuzzing consists of modifications to AFL as well as a neural network model to predict optimal locations to fuzz.

\subsection{Augmented-AFL}
We augmented the AFL fuzzer for this work to utilize the neural models. The Augmented-AFL queries a neural network model with each seed prior to fuzzing. The neural model categorizes the seed into useful and useless sections at the byte granularity, which is used during fuzzing. The mutations that target no useful sections are vetoed prior to execution. This augmented approach is depicted in Figure \ref{augmented_afl_plant}.

The AFL fuzzing strategy applies the following small localized mutations. Note that all mutations below are performed on sequential sections.

\begin{itemize}
\item Bit flips: Mutate the input by flipping [1/2/4] bit(s) at a time.
\item Byte flips: Mutate the input by applying exclusve or to [1/2/4] byte(s) with 0x$FF$.
\item Arithmetic mutations: Mutate the input by adding/subtracting interesting quantities at [1/2/4] byte granularities.
\item Interesting substitutions: Mutate the input by splicing an "interesting" value at [1/2/4] byte granularities.
\item Dictionary substitutions: Mutate the input by replacing bytes with user supplied "interesting" values. These may be longer than 4 bytes in length.
\end{itemize}

All mutations above are small and localized changes of which there are finitely many for a given seed. After the conclusion of the deterministic phase, AFL begins stacking many of these small localized mutations which are non-local and of significant hamming distance with respect to the input. AFL may apply somewhere between 2 and 128 stacking changes chosen uniformly. In addition to the previously mentioned location mutations, the following mutations may also be applied:

\begin{itemize}
\item Random byte assignment: Assign a random value to a random byte.
\item Delete bytes: Delete a section of the input file.
\item Clone bytes: Append bytes to a section of the input file.
\item Overwrite bytes: Overwrite a section of the input file.
\end{itemize}

\begin{figure*}
\removelatexerror
\centering
\begin{minipage}[t]{8cm}
  \vspace{0pt}
\begin{algorithm}[H]
\DontPrintSemicolon
\TitleOfAlgo{AFL Fuzzing}
\SetKwFunction{RandInt}{RandInt}
\SetKwFunction{len}{len}
\SetKwFunction{mutate}{mutate}
\SetKwFunction{Execute}{Execute}
\SetKwFunction{HasInputGain}{HasInputGain}

\KwInput{$Seeds$, Target program $P$}
  \KwResult{$MaliciousInputs$}
 \For{$Seed$ $\in$ $Seeds$}
{
 \HiLi \;
 \For{$iterations \gets 0$ \KwTo $limit$ }{
    $input \gets Seed$

    $length \gets$ \len{$Seed$}

    $mutations \gets$ \RandInt{$length$}

    \For{$mut \gets 0$ \KwTo $mutations$}
    {
        $loc \gets$ \RandInt{$length$}

        \mutate{$input$, $loc$}

    }
   \HiLi \;
\HiLi \;
\HiLi \;
\;
   $result, codecov \gets$ \Execute{$P$, $input$}

   \If {$result$ is crash}
   {
     Append $input$ to $MaliciousInputs$
   }
  \If {\HasInputGain{codecov}}
   {
   Append $input$ to $Seeds$
   }
 }
}
\end{algorithm}

\end{minipage}%
\begin{minipage}[t]{8cm}
  \vspace{0pt}
\begin{algorithm}[H]
\DontPrintSemicolon
\TitleOfAlgo{Augmented-AFL Fuzzing}
\SetKwFunction{RandInt}{RandInt}
\SetKwFunction{len}{len}
\SetKwFunction{mutate}{mutate}
\SetKwFunction{Execute}{Execute}
\SetKwFunction{HasInputGain}{HasInputGain}
\SetKwFunction{QueryModel}{QueryModel}
\SetKwFunction{Sum}{Sum}

\KwInput{$Seeds$, Target program $P$}
  \KwResult{$MaliciousInputs$}
 \For{$Seed$ $\in$ $Seeds$}
{
 \HiLi $bytemask \gets$ \QueryModel{$Seed$}

 \For{$iterations \gets 0$ \KwTo $limit$ }{
    $input \gets Seed$

    $length \gets$ \len{$Seed$}

    $mutations \gets$ \RandInt{$length$}

    \For{$mut \gets 0$ \KwTo $mutations$}
    {
        $loc \gets$ \RandInt{$length$}

        \mutate{$input$, $loc$}

    }
   \HiLi $\mathit{diff}$ $\gets input \oplus Seed$

   \HiLi \If{$\sum \mathit{diff}$ $\land$ $bytemask$ < $\alpha$}
   {
   \HiLi   \Continue
   }
   $result, codecov \gets$ \Execute{$P$, $input$}

   \If {$result$ is crash}
   {
     Append $input$ to $MaliciousInputs$
   }
  \If {\HasInputGain{codecov}}
   {
   Append $input$ to $Seeds$
   }
 }
}
\end{algorithm}

\end{minipage}
\caption{Comparison of the AFL-Augmented algorithm with the benchmark AFL algorithm. The primary difference is in using the QueryModel function to gain the bytemask of useful and useless locations. This bytemask is bitwise and'ed with the mutation diff to approve or reject the mutation prior to execution.}
\label{afl-augmented-comp}

\end{figure*}
Due to the location and context insensitive nature of AFL fuzzing, most mutations yield no input gain. The goal of augmented fuzzing is to improve the hit-rate of mutations. Using the annotated seed provided by the model, mutations that are unlikely to give input gain are avoided. We used a highly permissive veto approach to reject mutations which target no useful locations. The augmented mutation algorithm is presented in Figure ~\ref{afl-augmented-comp}.

\subsection{Neural network architectures}
We now describe different neural network architectures that we use to learn the coverage heatmap prediction function. Recall that the family of functions to be learnt is of the following format:

\begin{equation}
\left \{f_k: \left \{\text{0x}00, \text{0x}01, \ldots , \text{0x}FF \right \}^k \mapsto [0,1]^k \middle | k \in \mathbb{N} \right \}
\end{equation}

A possible encoding scheme for this family of functions is to feed the input "as-is" to an underlying neural network. This would involve encoding the data as a sequence of real valued floating point numbers in the range $[0, 255]$. However, this is suboptimal since the binary data does not necessarily represent \emph{magnitudes}, but could also represent \emph{states}. It is incorrect to assume that each byte represents a numerical quantity, it could represent bitmasks or other non-numerical values. We, therefore, encode the byte level information in the "sequence-of-bits" format:

\begin{equation}
\left \{f'_k: \left \{0,1 \right \}^{8k} \mapsto \left [0,1 \right ]^{8k} \middle | k \in \mathbb{N} \right \}
\end{equation}

This function determines usefulness at a bit granularity. We reconstitute $f$ given $f'$ by averaging the constituent bit values for each byte.

Due to the varying length and sequential nature of the input, Recurrent Neural Network (RNN) was the obvious choice. Each input file is sequential data which is most likely parsed sequentially by the target program. RNNs are able to count~\cite{rodriguez99}. This is useful in annotating file formats which contain header information at fixed offsets. RNNs have been successfully used in Statistical Machine Translation~\cite{cho14,bahdanau14}, and this task is similar since binary file formats can be considered a language.

RNNs are known to have problems with longer sequences. Due to this reason we chose Long Short-Term Memory (LSTM) as our base recurrent unit~\cite{hochreiter97}. LSTMs extend the memory capability of a recurrent unit to longer sequences. This is accomplished by a separate pathway for the flow of memory. LSTMs can also "forget" memories that have outlived their usefulness which allows more expressiveness for longer sequences. Recall that a recurrent unit computes state update and output, $h_t, o_t$ as follows.

\begin{equation}
h_t, o_t  = f(x_t, h_{t-1})
\end{equation}

An LSTM decomposes the above overall framework into the following subcomponents.
\begin{align*}
f_t &= \sigma(W_f \cdot [x_t, h_{t-1}] + b_f) \\
i_t &= \sigma(W_i \cdot [x_t, h_{t-1}] + b_i) \\
C_t &= f_t \times C_{t-1} + i_t \times tanh(W_C \cdot [x_t, h_{t-1}] + b_C) \\
o_t &= \sigma(W_o \cdot [x_t, h_{t-1}] + b_o) \\
h_t &= o_t \times tanh(C_t)
\end{align*}

where:
\begin{conditions}
\sigma  & Sigmoid activation function \\
W_{*} & Learned weight vectors \\
b_{*} & Learned bias vectors \\
\end{conditions}

\begin{table*}[ht]
\centering
\caption{Enumeration of all architectures, with their number of trainable parameters. Blank entries mean this combination was not tested.}
\label{my-label}
\begin{tabular}{|l|l|l|l|l|}
\hline
 & \multicolumn{2}{|c|}{\bf One Layer} & \multicolumn{2}{|c|}{\bf Two Layers}  \\ \hline
{\bf Model} &  {\bf 64-bit chunk} & {\bf 128-bit chunk} & {\bf 64-bit chunk} & {\bf 128-bit chunk} \\ \hline \hline
LSTM & 33,024  & 131,584  & 66,048  & 263,168 \\ \hline
Bidirectional LSTM & 66,048 &  263,168 & 57,856 & 230,400 \\ \hline
Seq2Seq & N/A & N/A & 57,856 & 230,400 \\ \hline
Seq2Seq+Attn & N/A & N/A & 57,985 & 230,657  \\ \hline
\end{tabular}
\label{param_grid}
\end{table*}

\begin{figure*}[t]
\begin{center}
\includegraphics[scale=1.2]{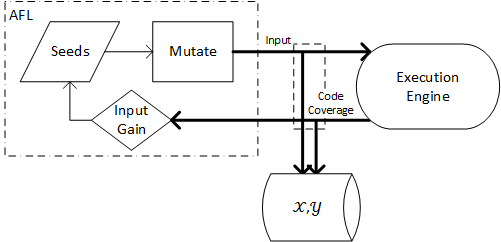}
\end{center}
\caption{The data collection approach to collect input, code coverage pairs.}
\label{data_collection}
\end{figure*}

The forget gate $f_t$, and the input gate $i_t$ control whether old memory is forgotten, and whether the current input is worth remembering. This interplay allows the memory information of LSTMs to persist through longer sequences.

We explored several architectures using LSTM as our base recurrent unit to determine whether there is an affinity between input annotation and neural architecture. Recent advances in Neural Machine Translation have highlighted some important architectures such as Seq2Seq ~\cite{cho14} and Seq2Seq with Attention ~\cite{bahdanau14}. Do these translation focused architectures also work well when learning the structure of the binary formats? In total, we evaluated the following architectures:

\begin{itemize}
\item LSTM: A simple unidirectional LSTM~\cite{hochreiter97}.
\item Bidirectional LSTM: A bidirectional LSTM which sees the input backwards and forwards.
\item Seq2Seq: Sequence-to-sequence architecture~\cite{cho14}.
\item Seq2Seq+Attn: Sequence-to-sequence architecture with attention~\cite{bahdanau14}.
\end{itemize}

Bidirectional LSTMs see the input in backward and forward order. A bidirectional LSTM is composed of two unidirectional LSTMs, one in each forward and backward direction. Given a length $n$ sequence, to compute the values for timestep $t$, the forward LSTM's $h_{t-1}$ and the backward LSTM's $h_{n-t-1}$ are used in conjunction. A merge function is used to merge the output of the unidirectional LSTMs. The merge function can be one of many functions combining two like sized vectors such as sum, multiply, or concatenate. We chose to use the sum function for one-layer Bidirectional LSTMs, and the concatenate function for LSTMs with two or more layers.

We also experimented with the number of layers, and the chunk size of the LSTM inputs provided at each timestep. The purpose was to determine how \emph{complex} byte prediction is, and whether more complex models outperform simpler models.

Our models consume $k$ bits per iteration, and also output $k$ bits per iteration. We experimented with chunking the input sequence into 64-bit or 128-bit chunks. Our proposed architectures and total number of trainable parameters are detailed in Table \ref{param_grid}.

One-layer bidirectional LSTMs used a sum merge function, while two-layer bidirectional LSTM used a concatenate function. The second layer of the two-layer bidirectional LSTM was a unidirectional LSTM. Seq2Seq and Seq2Seq+Attn were comprised of one encoding, and one decoding layer. The encoding layer was a bidirectional LSTM which was merged using the concatenate function. The decoding layer was a unidirectional LSTM. We did not explore unidirectional Seq2Seq or Seq2Seq+Attn.

\section{Evaluation}

We evaluate the effectiveness of the Augmented-AFL on four target programs with the goal of assessing the augmentation strategy across a diverse set of programs encountered in practice. The chosen target programs were readpng~\cite{libpng}, readelf~\cite{binutils}, mupdf~\cite{mupdf}, and libxml~\cite{libxml}. We investigated several metrics for these programs, primary amongst them were code coverage and input gain. Code coverage and input gain are first class metrics used by AFL. Input gain is measured by the total number of inputs which cause input gain over the runtime of the fuzzer. The number of crashes found with Augmented-AFL and AFL were also measured.

\begin{table*}[t]
\centering
\caption{Enumeration of all programs, all architectures reporting on average code coverage after a 24 hour evaluation run. Top 3 performing strategies are bolded for each program.}
\label{my-label2}
\begin{tabular}{|l|l|l l l l|}
\hline
& & \multicolumn{2}{|c|}{One Layer} & \multicolumn{2}{|c|}{Two Layer}  \\ \hline
& Model &  64-bit chunk & 128-bit chunk & 64-bit chunk & 128-bit chunk \\ \hline \hline
\multirow{5}{*}{readelf} & AFL Benchmark & \multicolumn{4}{l|}{12.20\%} \\
& LSTM & \textbf{13.46\%}  & \textbf{13.34\%}  & 12.83\%  & 12.93\% \\
& Bidirectional LSTM & 12.01\% &  12.75\% & 12.50\% & 12.94\% \\
& Seq2Seq & N/A & N/A & 12.11\% & \textbf{13.10\%} \\
& Seq2Seq+Attn & N/A & N/A &12.59\%   & 12.90\%   \\ \hline \hline
\multirow{5}{*}{readpng} & AFL Benchmark & \multicolumn{4}{l|}{2.20\%} \\
& LSTM & \textbf{2.43\%}  & 2.38\%  & 2.35\%  & 2.37\% \\
& Bidirectional LSTM & 2.39\% &  2.39\% & \textbf{2.42\%} & 2.38\% \\
& Seq2Seq & N/A & N/A & 2.31\% & 2.31\% \\
& Seq2Seq+Attn & N/A & N/A & \textbf{2.42\%}  & 2.39\%  \\ \hline \hline
\multirow{5}{*}{mupdf} & AFL Benchmark & \multicolumn{4}{l|}{11.63\%} \\
& LSTM & 11.44\%  & 11.57\% & 11.26\%  & 11.17\% \\
& Bidirectional LSTM & \textbf{11.80\%} & 11.48\% & 11.61\% & 11.39\% \\
& Seq2Seq & N/A & N/A  &  11.56\% & 11.58\% \\
& Seq2Seq+Attn & N/A & N/A & \textbf{11.71\%} & \textbf{11.65\%}  \\ \hline \hline
\multirow{5}{*}{libxml} & AFL Benchmark & \multicolumn{4}{l|}{2.09\%} \\
& LSTM & \textbf{2.11\%}  & 2.10\%  & \textbf{2.11\%}  & 2.10\% \\
& Bidirectional LSTM & 2.09\% & 2.11\% & \textbf{2.11\%} & 2.11\% \\
& Seq2Seq & N/A & N/A & 2.09\% & 2.11\% \\
& Seq2Seq+Attn &  N/A & N/A & 2.11\% & 2.11\% \\ \hline
\end{tabular}
\label{ccov_grid}
\end{table*}

We collected $180$ randomly chosen seed files for each program from a large sample population. The seed files were evenly divided into a training and test set. To collect the data for training the models, AFL was run for 24 hours. Input, code coverage pairs were collected at a uniform $1\%$ sampling rate. This collection strategy is highlighted in Figure \ref{data_collection}. Prior to training, the data was filtered with the strictly less than function with a cutoff of 0 to form the training set. That is, given a set of $(x,x', b, b')$, a training dataset ${\cal{X}}{\cal{Y}}$ is constructed as follows.

\begin{align}
{\cal{X}}{\cal{Y}} \> = \> \left \{ \left (x, x \oplus x' \right ) \> \middle | \sum_{i} \left [b_i < b'_i \right ]  > 0 \right \}
\end{align}

The model implementations were designed using Keras~\cite{chollet15}, a high level deep learning library. We chose to use Tensorflow~\cite{tensorflow15} as the low level backend for Keras.

The training data was heterogenous in length, and could consist of very large input files up to 200 Kilobytes in length. To mitigate these issues, input data longer than $10 kB$ was segmented into a set of $10 kB$ segments. After segmenting, the data was binned according to length and padded to the nearest chunk sized boundary. Each step of training consisted of selecting a bin proportional to bin size, and constructing a minibatch of elements in the selected bin. The models were trained for 12 hours to ensure convergence and the training was performed on Nvidia K40M GPUs with 12 gigabytes of RAM. We used a loss function of mean absolute error (MAE) and used the Adam optimizer~\cite{adam} with a learning rate of $5 \times 10^{-5}$ to train the model.

Although Augmented-AFL can \emph{exploit} previously learned patterns to improve the hit-rate of mutations, it does not \emph{explore} as well as the benchmark AFL algorithm. To counteract this tendency, for each seed Augmented-AFL may choose with $50\%$ probability to utilize the unaugmented fuzzing strategy. This allows for a good mix between exploration and exploitation. There exist many techniques to better achieve balance between exploration and exploitation that we hope to pursue in future.

For evaluating the learnt models, we restart AFL and Augmented-AFL on the test set of seed files. This evaluation phase was run for 24 hours. To minimize variance, many instances of AFL were run at one time. For AFL, 16 instances of AFL were run on a 16 core machine. For Augmented-AFL, 8 instances of AFL were run on a 16 core machine, with 8 cores reserved for model querying. For the majority of validation we used Azure Standard F16s machines with Intel Xeon E5-2673 v3 CPUs at 2.40GHz and 32GB of RAM. Due to out of memory issues, Azure Standard D14 size VMs were used for a small minority of cases. The Azure Standard D14 VMs are identical to Standard F16s, except with 112GB of RAM. Dynamic CPU frequency scaling was not enabled during validation. After 24 hours of execution, the data was averaged over the many instances.

\begin{figure*}[t]
\begin{center}
\includegraphics[width=.45\textwidth]{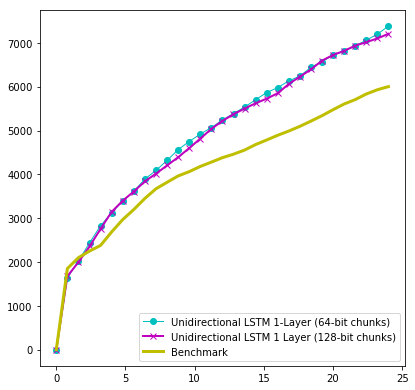}
\includegraphics[width=.45\textwidth]{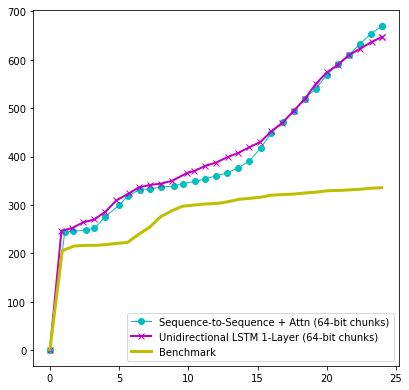}
\includegraphics[width=.45\textwidth]{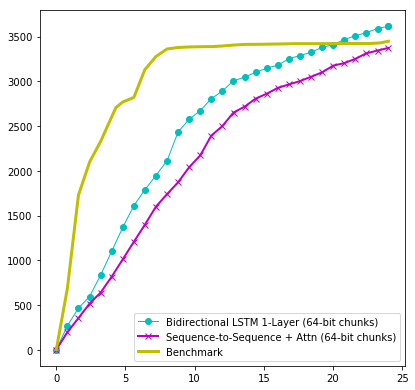}
\includegraphics[width=.45\textwidth]{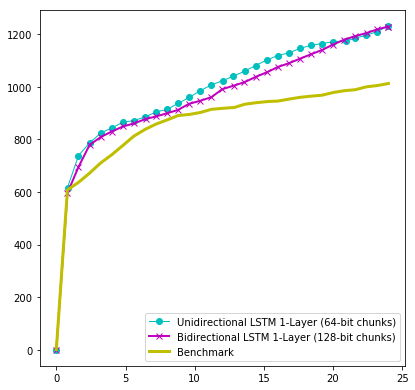}
\end{center}
\caption{Input gain vs time plots. From top left to bottom right: readelf, readpng, mupdf, libxml. X-axis denotes time since start in hours. Y-axis denotes unique number of code paths found.}
\label{afl-ig}
\end{figure*}

\subsection{Code Coverage}

The code coverage is reported for all architectures on all programs in Table \ref{ccov_grid}. We can observe significant improvements in the code coverage metric for readelf and readpng programs. Almost all models outperform the benchmark (baseline AFL) for these programs. Often times, the simplest unidirectional models outperform the other more complex models. However, no significant improvements in code coverage metric was observed with mupdf and libxml. For mupdf, most augmented models perform worse than the benchmark. The only exception is Seq2Seq+Attn model for mupdf that outperforms the benchmark by a small amount. For libxml, all models perform similarly concerning code coverage. The reported code coverage are all centered around $2.10\%$ and within the margin of error.

\subsection{Input Gain}
A second metric for measuring efficacy is input gain. Input gain is the number of paths found that exhibit never-seen-before behavior in the target program. This behavior is characterized by either executing a new block of code, or increasing the frequency of execution of a previously executed code block. Input gain vs time plots are presented for the two highest performing models for each program in Figure \ref{afl-ig}.

For all programs except for PDF, significant improvements in input gain are observed. This is expected for readpdf and readelf since code coverage typically increased for these programs. However, libxml did not exhibit code coverage increases during validation. This likely means that the same code sections were exercised more thoroughly with a variation in execution frequency.

The mupdf parser did not exhibit significant improvements in code coverage or input gain. We believe this shows the model query vs execution tradeoff in the proposed design. Because typical PDF files are quite large in size (over $100 kB$), the model query time adversely affects the performance of Augmented-AFL. Prior to fuzzing each seed file, the model must be \emph{queried} for this seed file, which can be several seconds for such large seed files. Over the runtime of the fuzzer, this query time adversely affects total fuzzing performance because execution is often blocked on the model query. We believe that throughput and performance improvements to model querying will improve the efficacy of the Augmented-AFL technique for lengthy formats such as PDF.

\begin{figure*}[t]
\begin{center}
\includegraphics[width=.45\textwidth]{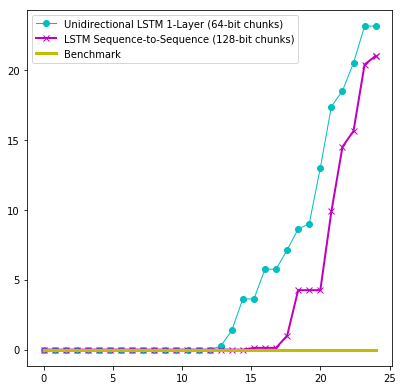}
\includegraphics[width=.45\textwidth]{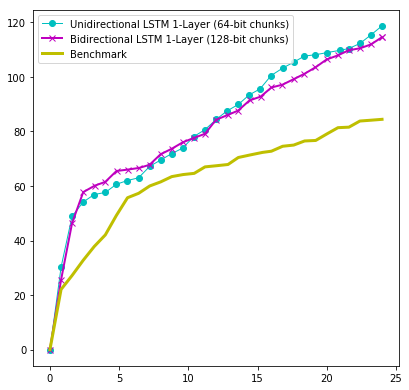}
\end{center}
\caption{Unique crashes vs time plots. Left to right: readelf, libxml. X-axis denotes time since start in hours. Y-axis denotes number of unique crashes found. }
\label{afl-crashes}
\end{figure*}

\subsection{Crashes}
The most important metric for measuring fuzzing efficacy is the number of malicious inputs discovered, which we measure by recording the number of unique crashes discovered during the execution. We observed crashes for only readelf and libxml and we therefore omit the plots for readpng and mupdf. The plots of unique crashes found over time are presented for readelf, and libxml in Figure \ref{afl-crashes}.

The augmented-AFL outperforms the AFL benchmark for both programs. For readelf, several unique crashes (more than 20) are observed by the 24-hour mark, whereas no crashes are observed for the benchmark. Similarly, the augmented-AFL results in finding about 110 unique crashes for libxml by the 24-hour mark in comparison to AFL discovering about 80 unique crashes. These results show  a significant improvement over the baseline.

These results show improvements over AFL, a state-of-the-art genetic algorithm greybox fuzzer. Using machine learning to predict interesting locations to fuzz, we are able to improve fuzzing performance over a wide variety of target programs and file formats. We, therefore, believe that machine learning guided fuzzing is a promising technique to improve greybox and blacbkox fuzzers, and similar techniques can be applied to learn several other fuzzing parameters in future.

\section{Related Work}

We now briefly discuss some of the related work on using machine learning techniques (in particular neural network based models) for guiding program fuzzing and program analysis.

\paragraph{Machine Learning for Grammar-based Fuzzing} The Learn\&Fuzz~\cite{learnfuzz} approach was recently developed for training neural networks (LSTMs) to learn generative models of the input formats for grammar-based fuzzing. For complex input formats such as PDF, randomly mutating the inputs would quickly result in invalid inputs, and therefore typically grammar-based fuzzing techniques are used to define the input grammars for such formats. However, writing a grammar manually by hand is tedious and error-prone especially for complex input formats. Learn\&Fuzz presented a technique to use LSTMs to learn a grammar (distribution) for PDF objects using a character-level model, which can then be sampled to generate new inputs. Instead of learning a grammar, our technique uses neural networks to learn a function to predict promising locations in a seed file to perform mutations. We believe our technique can complement Learn\&Fuzz to further improve neural grammar-based fuzzing.

\paragraph{Bandit formulation for Fuzzing}
Our work was chiefly inspired through bandit optimization techniques. There has been some work combining fuzzing and bandit optimization with fuzz configuration scheduling~\cite{woo13}. In particular, Woo et al.~\cite{woo13} modeled fuzzer configuration options as a bandit problem. However, our work takes this approach further by modeling fuzzing as a bandit problem. We believe that fuzzing is a discrete optimization problem which can be simplified by identifying the subset of byte locations with the most payoff, and that the identification of optimal byte locations is a problem best solved through multi-armed bandits approach. This ``bytes-as-bandits" approach deserves further study, in particular, we hope to shed further light on theoretically optimal methods of identifying optimal bytes.

\paragraph{Evolutionary Fuzzing}
Evolutionary fuzzing uses execution feedback to guide future mutation decisions. Some early work along this direction includes the Evolutionary Fuzzing System (EFS) by DeMott et al.~\cite{demott07}. EFS uses Genetic Algorithm techniques to evolve the seed pool where the fitness function is defined as the induced code coverage. EFS uses several sophisticated crossover methods to evolve the seed pool over time. In contrast to AFL, EFS only uses gene crossover methods to "fuzz" the set of seed files. Recent advances in evolutionary fuzzing include Taint-based Directed Whitebox Fuzzing~\cite{ganesh09} and VUzzer~\cite{rawat17}. Taint-based Directed Whitebox Fuzzing uses dynamic taint tracing to identify parts of the seed which may cause ~\emph{dangerous} code sections to be executed. Mutations are \emph{directed} towards these sections to discover bugs. VUzzer takes a similar approach of using dynamic taint tracing, but does not attempt to identify and focus on dangerous code sections. VUzzer works towards increasing code coverage and thoroughly exercising the code.

The common theme amongst the above techniques is a feedback loop dependent on past execution behavior. Although our approach also contains a feedback loop, we favor a neural approach to guide future fuzzing actions. This is novel because of the ease of development and integration. Our neural guided approach can be developed quickly using off the shelf Deep Learning libaries, and can be easily integrated into an existing Greybox or Blackbox fuzzer. Our approach has relatively low overhead since the simple models have low query time and the coverage map can be computed efficiently.

\paragraph{Neural Networks for Program Analysis} There have been several recent works proposed for training neural networks to perform program analysis such as program repair~\cite{neuralrepair}, program optimization~\cite{neuralcompilation}, and program synthesis~\cite{neuralsynthesis,robustfill,ap}. These works learn neural representations of programs to perform various prediction tasks, whereas in our work we train the neural models to instead represent input files. Moreover, our work presents the first application of training neural networks to learn promising fuzzing locations in input files.

\section{Future Work and Conclusion}
We have demonstrated a novel neural based augmentation to greybox fuzzing. This augmentation identifies useful locations to fuzz in seed files. We believe that most binary file formats contain small sections which highly affect execution behavior of the program. Focusing fuzzing on these small sections is useful since they are likely to yield novel execution behavior in the target program.

Our augmentation was chiefly targeted towards greybox fuzzers such as AFL. Greybox fuzzers are the perfect testbed because they provide code coverage feedback for each execution. This feedback was used to train a neural network model to identify the most promising locations for fuzzing. Our approach is simple and easy to integrate with most greybox fuzzers.

We showed that recurrent models such as LSTM work well for this task. This task can be considered similar to Statistical Machine Translation. Recurrent models have had great success on statistical machine translation task in recent years. We evaluated the model on a variety of target binary file formats such as PDF, XML, PNG, and ELF. The model significantly outperformed the state-of-the-art AFL fuzzer on all target programs except PDF. Typically the simplest models outperformed the more complex models. We believe that the model performance on PDF shows the cost-benefit tradeoff of querying the model on large input files. However, we believe it is possible to improve the results on large file formats such as PDF with some additional performance engineering.

Although our results are promising, there are many avenues for further work. We trained our current models in an offline supervised setting. A natural extension to this work would be to make learning online using reinforcement learning such that the model continuously improves as the fuzzing process proceeds. We believe that fuzzing can be greatly enhanced through "feedback loop fuzzing," where past execution behavior guide future mutations. We envision a new type of fuzzer which leverages machine learning models to guide its mutation decisions. Fuzzing provides a treasure trove of high fidelity structured data. The signal to noise ratio is high. Another possible extension along this avenue is using generative models. Our model is restrictive, where mutations proposed by AFL are vetoed. A more interesting approach is to generate the mutations that are applied to seed files, which we plan to consider in near future.


\bibliographystyle{plain}
\bibliography{arxiv_2017}

\end{document}